# Microscopic piezoelectric behavior of clamped and membrane (001) PMN-30PT thin films

A. Brewer[1]†, S. Lindemann[1]†, B. Wang[2], W. Maeng[1], J. Frederick[1], F. Li[3], Y. Choi[4], P. J. Thompson[5], J. W. Kim[4], T. Mooney[4], V. Vaithyanathan[6], D. G. Schlom[6,7,8], M. S. Rzchowski[9], L. Q. Chen[2], P. J. Ryan[4,10]*, and C. B. Eom[1]*

[1]*Department of Materials Science and Engineering, University of Wisconsin-Madison, Madison, WI 53706, USA*
[2]*Department of Materials Science and Engineering, The Pennsylvania State University, University Park, PA 16802, USA*
3*Electronic Materials Research Laboratory, Key Laboratory of the Ministry of Education, Xi'an Jiaotong University, Xi'an, 710049, China*
[4]*X-Ray Science Division, Argonne National Laboratory, Argonne, IL 60439, USA*
[5]*Department of Physics, University of Liverpool, Liverpool, L69 3BX, UK*
[6]*Department of Materials Science and Engineering, Cornell University, Ithaca, NY 14853, USA*
[7]*Kavli Institute at Cornell for Nanoscale Science, Ithaca, NY 14853, USA*
[8]*Leibniz-Institut für Kristallzüchtung, Max-Born-Straße 2, 12489 Berlin, Germany*
[9]*Department of Physics, University of Wisconsin-Madison, Madison, WI 53706, USA*
[10]*School of Physical Sciences, Dublin City University – Dublin 9, Ireland*

**Bulk single-crystal relaxor-ferroelectrics, like $Pb(Mg_{1/3}Nb_{2/3})O_3$-$PbTiO_3$ (PMN-PT), are widely known for their large piezoelectricity. This is attributed to polarization rotation which is facilitated by the presence of various crystal symmetries for compositions near a morphotropic phase boundary (MPB). Relaxor-ferroelectric thin films, which are necessary for low-voltage applications, suffer a reduction in their piezoelectric response due to clamping by the passive substrate. To understand the microscopic behavior of this adverse phenomenon, we employ AC electric field driven *in-operando* synchrotron x-ray diffraction (XRD) on patterned device structures to investigate the piezoelectric domain behavior under an electric field for both a clamped (001) PMN-PT thin film on Si and a (001) PMN-PT membrane released from its substrate. In the clamped film, the substrate inhibits the field-induced rhombohedral (R) to tetragonal (T) phase transition resulting in a reversible R to Monoclinic (M) transition with a reduced longitudinal piezoelectric coefficient $d_{33}$ < 100 pm/V. Releasing the film from the substrate results in recovery of the R to T transition and results in a $d_{33}$ > 1000 pm/V. Using diffraction with spatial mapping, we find that lateral constraints imposed by the boundary between active and inactive material also inhibits the R to T transition. Phase-field calculations on both clamped and released PMN-PT thin films simulate our experimental findings. Resolving the suppression of thin film piezoelectric response is critical to their application in piezo-driven technologies.**

† These authors contributed equally
*correspondence should be addressed: eom@engr.wisc.edu, pryan@aps.anl.gov

Relaxor-ferroelectrics are of particular interest due to their strong piezoelectric responses in bulk single crystals when their compositions lie near a morphotropic phase boundary (MPB).[1] Rhombohedral (1-x)Pb(Mg$_{1/3}$Nb$_{2/3}$)O$_3$-(x)PbTiO$_3$ (PMN-xPT) single crystals demonstrate particularly high piezoelectricity with a $d_{33} > 1500$ pm/V when applying the field along one of the <001> directions.[1] This response arises from a reversible rotation of the spontaneous polarization to align with the applied electric field, causing a change in symmetry from rhombohedral (R) to the tetragonal (T) phase. This polarization rotation occurs through one or more monoclinic variants ($M_a$, $M_b$, or $M_c$)[2-7], which are highlighted in **Fig. 1(a)**. The monoclinic symmetries are seen as a key bridging mechanism between the morphotropic phases, allowing for the flexibility of polarization rotation pathways.[8,9]

Thin films of ferroelectrics suffer a suppression of their piezoelectric effect due to mechanical and electrical boundary conditions. In the patterned electrode thin-film device geometry, mechanical clamping by the substrate reduces the piezoelectric response by resisting elastic deformation[10] or even changes the phase distribution with composition[11]. Additionally, the extrinsic piezoelectric effect from domain wall motion is also dampened due to lateral lattice restrictions.[12] Unfortunately, the microscopic mechanisms of substrate clamping on spontaneous polarization rotation are insufficiently understood.

In this letter we show a recovery of giant piezoelectricity in single crystal PMN-PT thin films via complete removal of the substrate, and consequently, the associated mechanical clamping. Dynamic synchrotron x-ray diffraction (XRD) on patterned device structures with applied AC electric field and phase-field simulations are used to investigate both clamped (001)-PMN-30PT films on Si (**Fig. 1(b)**) (i.e. (1-x)PMN-(x)PT with x = 30% and (001) crystal axis perpendicular to the plane of the film) as well as (001)-PMN-30PT membranes removed from their substrate (**Fig. 1(c)**). Sample fabrication details and XRD methodology can be found in **Supplementary Notes 1 and 2** of the supplemental section. The clamping by the substrate results in a reduced piezoelectric effect with a reversible phase transition between R and $M_a$. Removed from its substrate, the PMN-PT membrane shows a complete polarization rotation from R to T symmetries which results in the recovery of giant piezoelectricity. In the membrane, we develop a "ratcheting" process using minor PE-loops that nucleates and grows the emergent M and T phases. Thus, we observe the R to T transition through the morphotropic boundary regime in a stepwise manner, allowing us to determine the polarization rotation pathway in PMN-PT single crystal thin films. Spatially dependent XRD demonstrated how the lateral boundary between active and inactive material additionally inhibits the piezoelectric response. Our work demonstrates that complete removal of substrate clamping (replaced by the much softer interaction with PDMS), and careful consideration of lateral boundary conditions are necessary to maximize piezoelectricity in PMN-PT thin films.

**Figure 1(d)** shows the polarization vs electric field (PE) hysteresis for the clamped and membrane PMN-PT films taken at a 10 kHz frequency. **Fig. S1** shows the same PE loop for the PMN-PT membrane (**Fig. S1(a)**), as well as a 1 kHz PE loop for the clamped PMN-PT film used in the XRD experiments (**Fig. S1b**). The positive electric field points from the SrRuO$_3$ (SRO) electrode towards the Pt electrode. The ferroelectric (FE) imprint possibly arises from the asymmetric electrode configuration[17] and corresponds to the polarization states of the PMN-PT pointing towards the SRO electrode at zero bias. Double-beam laser interferometry (DBLI) in **Fig. 1e** indicates the clamped film exhibits a maximum longitudinal piezoelectric coefficient ($d_{33}$) of ~30 pm/V while the membrane shows a larger $d_{33}$ of ~1100 pm/V, illustrating the recovery of giant piezoelectricity in the PMN-PT membrane.

L-scans of the 004 peak were performed under 1 kHz AC triangular waveform sweeps. **Fig. S2(a)** shows data from the center of the film pattern. Data at various locations within the pattern gave identical results for the clamped film. The low-field sweep resulted in a shift of the 004 PMN-PT peak position, indicating a change in lattice parameter, while the high-field sweep resulted in the peak shifting position and changing shape, suggesting the presence of a second reflection. Subsequent peak fitting was performed with the best fit corresponding to two phases which we identified as R and $M_a$ in comparison to bulk lattice parameters taking into consideration in-plane strain.[15] Representative scans with peak fitting are shown in **Fig. S2(b)**. With 150 kV/cm, **Fig. 2(a)** shows the volume percentage of R and $M_a$ phases (calculated from the relative areas under the curves of the fitted peaks) vs E-field, and **Fig. 2(b)** shows the lattice parameters vs E field for the R and $M_a$ peaks. **Figure S3** shows the volume percentage and lattice parameters for 50, 100, and 150 kV/cm sweeps.

At zero bias, the film consists of ~85% majority R phase and ~15% minority $M_a$ phase (**Fig. 2(a)**). The 50 kV/cm response does not show significant exchange between R and $M_a$ except for a small volume at -50 kV/cm, (**Fig. S3(a)**). As a result, only an E-field dependent linear response of the R lattice parameter was observed (**Fig. S3(b)**), resulting in a $d_{33}$ of only 69 pm/V (**Fig. S4**). Under the 100 kV/cm sweeping field (**Figs. S3(c) and S3(d)**), the R lattice parameter shows butterfly-loop switching behavior while there is only exchange between R and $M_a$ volumes at the negative bias due to the FE imprint. With the application of the larger 150 kV/cm AC field, **Fig. 2(a)** shows significant exchange between R and $M_a$ phase volumes only, indicating that the tip of the polarization vector traverses the crystal face diagonal as indicated in **Fig. 1(a)**. The behavior of the lattice parameters (**Fig. 2(b)**) indicates strong elastic coupling between the two phases. As the applied bias increases away from the FE imprint at 20 kV/cm, the R phase remains the majority phase and exhibits lattice expansion while the $M_a$ phase shows anomalous contraction. Around -75 kV/cm and 100 kV/cm the lattice response of both phases show a reversal in behavior (**Fig. 2(b)**). As $M_a$ becomes the primary phase driving the piezoresponse, the R volume now mitigates the $M_a$ imposed lateral strain showing anomalous contraction with increasing field. Lateral clamping by the substrate generates this competitive coupling between R and $M_a$ phases and reduces the overall response of the film. Upon removal of the bias the system returns to the initial phase volume ratio indicating full reversibility. In bulk, the R phase does not recover but remains in the monoclinic variant $M_a$ below 350 K.[5]

With the Si substrate removed, L scans around the 004 reflection observe the piezoresponse in the PMN-PT membrane without the substrate clamping effect. To study the effects of constraints imposed by neighboring unbiased film, we measured various locations across the device (**Fig. 3(a)**). **Figure 3(b)** shows XRD intensity (indicated by the color-bar) vs lattice parameter, measured under static conditions (zero applied field) after applying a 1 kHz AC bias of 50 kV/cm for a series of time increments of 120 s, 480 s, and 520 s. An initial spatially resolved scan was performed before any electric field was applied to the sample to measure the as-grown (virgin) state of the PMN-PT film (**Fig. 3(b)**). Slices of the XRD intensity, indicated by white dashed lines, are plotted below for three representative locations in the electrode, including the fitting results of R, M, and T phases.

For the virgin electrode state of the membrane, the PMN-30PT structure was uniform across the entire electrode and consisted of 90% R phase with a pseudocubic lattice parameter of 4.013 Å, and 10% $M_a$ phase with pseudocubic lattice parameter 4.029 Å. After 120 s of AC bias (**Fig. 3(b)**) the volume of the rhombohedral phase decreases and the monoclinic phase volume increases in the center of the electrode. An additional phase appears with a lattice parameter of 4.065 Å, which is identified as tetragonal (T) based on a comparison with the tetragonal phase in unbiased bulk PMN-xPT above x = 37%.[16] The larger volume fractions transitioning from R to M and M to T toward the center of the electrode (-30 μm), compared with the edge of the electrode (-110 μm), indicates that clamping between biased and unbiased PMN-PT regions can decrease the piezoresponse by inhibiting phase exchange. After 480 seconds the M and T phases at the center have a larger out-of-plane lattice parameter than near the edge. This indicates that the unbiased piezoelectric material around the electrode not only inhibits the phase transition, but also limits the lattice expansion of the transitioned volumes. This is most evident in the last measurement, performed after 520 seconds of applied AC bias. Lines have been added to the slices of the 520 seconds data as a guide to see that the lattice parameter of the R phase remains unchanged from the edge to the center of the electrode while the lattice parameters of the M and T phases increase toward the center. The lattice parameters of the virgin and 520 s states versus position are plotted in **Fig. 3(c).** For the virgin state there is no position dependence of the R and $M_a$ lattice parameters while the T and M phases present after biasing for 520 s exhibit greater lattice expansion further from the inactive material. Here M includes all three monoclinic phases ($M_a$, $M_b$, and $M_c$).

**Figure 4** elucidates the polarization rotation pathway in the membrane by detailed analysis of the zero-field XRD results after 480 s AC biasing (**Fig. 4a**) and fits of the XRD data to fractions of the R,T, and all monoclinic variants for various locations across the electrode (**Fig. 4(b)**). **Figure 4(c)** summarizes the resulting volume percentage of each phase at locations across the membrane. The near-edge region (-150 µm) has mostly maintained the R phase predominant throughout the entire membrane in the virgin state, with the second most prevalent phase being $M_b$. Moving toward the center of the membrane, the R and $M_b$ phase volumes decrease, while T and $M_c$ phase volumes increase. The $M_a$ phase volume is roughly constant across the membrane. The involvement of both the $M_b$ and $M_c$ phases in the membrane contrasts the clamped film behavior where the polarization was constrained to the $M_a$ plane. Without clamping, the spontaneous polarization is no longer restricted along $M_a$ and can transition through the $M_{b,c}$ phases, allowing for easier volume exchange between R and T. No T domains were observed in the clamped film in this electric field regime.

**Figure 5** shows phase-field simulations[18] (details in **Supplementary Note 3**) replicating a multi-domain system for both clamped and membrane PMN-PT films as their phase distribution evolves under an applied time-independent electric field. These simulations did not accurately distinguish $M_a$, $M_b$, and $M_c$ phases. For the clamped film, the R phase initially dominates with a small amount of M giving rise to the initial multi-domain mesoscopic structure shown in **Fig. 5(a)**. As an electric field is applied, the M volume grows at the expense of the R phase, reaching an evenly mixed configuration at ~50 kV/cm. By 100 kV/cm, the entire film becomes M. Upon removing the applied bias the film relaxes back to an equilibrium state that is primarily R phase, in agreement with experiment.

As seen in XRD, the phase composition of the PMN-PT membrane does not change upon removal of the substrate. In the simulation, the initial configuration of the membrane is also similar to the clamped film, as shown in **Fig. 5(b)**. With electric field, the entire film becomes M phase at ~10 kV/cm, a much lower field than what was required for the clamped film, indicating the significant role of substrate clamping in stabilizing the R and M phases. Upon increasing the field to 25 kV/cm, the entire membrane becomes T phase. Once the field is removed, the T phase remains stable as observed by XRD (**Fig. 3(b)**). In bulk PMN-PT, the T phase reverses back to M phase upon field removal. Nonetheless, the presence of the FE imprint in the PMN-PT membrane stabilizes the T phase (**Supplemental Fig. S5**). The simulation results match closely with the XRD for the membrane in the center of the electrode area, far from the boundaries, as the simulations have not taken into account the boundary conditions between active and inactive PMN-PT regions.

The XRD studies of the clamped and membrane PMN-30PT films provide crucial information about how substrate clamping reduces the piezoresponse of thin films. As seen in **Fig. 1(e)**, $d_{33}$ of the clamped film is on the order of ~30 pm/V (red curve) while the membrane is over 1000 pm/V (blue curve) and is comparable to that of bulk PMN-PT. XRD of the clamped film showed that even with application of a 150 kV/cm AC bias, only a small rotation of the spontaneous polarization occurs as R volumes exchange with $M_a$ (**Fig. 2(a)**), but no amount of T phase is observed. Interaction between R and $M_a$ results in conflicting lattice expansion/contraction behavior, further limiting the longitudinal piezoelectric response.

In the PMN-PT membrane, rotation of the spontaneous polarization between R and T symmetries is easily achieved. The substrate clamping restricted the polarization to the $M_a$ plane, but with the substrate removed the polarization was able to follow its preferred pathway along $M_{b,c}$. Due to the ferroelectric imprint (**Fig. S1(a)**), each minor loop initiates nucleation of the monoclinic and tetragonal phases and allows them to grow through irreversible domain wall motion without providing a driving force to rotate their polarization back to R. This accumulated 'ratchet' process drives a systematic 'one-direction' evolution through the MPB transition. We argue that the number of AC cycles at fixed maximum electric field acts analogously to a variable DC electric field. The structural evolution with cycle number thus represents the polarization pathway. As seen in **Fig. 4(c)**, moving from the edge to the center of the electrode, the R phase becomes the minor phase as it exchanges volume with the M and T phases. Even as the M and T phases increase their lattice parameters, the R phase maintains a constant out-of-plane lattice parameter as shown in **Fig. 3(c)**. Freed from the substrate clamping constraints, the minor phases no longer exhibit anomalous lattice contraction. Based on the R and T lattice parameters measured here, a theoretical estimation of the PMN-PT membrane's $d_{33}$ (assuming 100% conversion of R domains to T domains under 5V bias) would result in a large $d_{33}$ of nearly 1300 pm/V (see **Supplementary Note 4**), which is consistent with our DBLI results (**Fig. 1(e)**), nearing values observed in bulk PMN-PT. The DBLI actually shows a $d_{33}$ of ~1000 pm/V, likely as a result of less than 100% conversion of R to T domains as observed in our XRD.

Lastly, our measurements of the spatially dependent response of the PMN-PT membrane to an electric field show that the surrounding unbiased material can further inhibit the piezoresponse of the film (**Figs. 3 and 4**). The largest piezoelectric effect is seen at the center of the electrode, exhibiting both phase transition from R to T while also experiencing lattice expansion. The edges of the electrode, however, show reduced piezoelectric response through maintaining a majority of R domains while also showing less lattice expansion due to lateral clamping with unbiased material which persists up to 100 um into the electrode area. Further maximizing the piezoelectric coefficients in the PMN-PT membrane may be achieved by removing unbiased material and creating PMN-PT island structures in order to allow for uninhibited rotation of the spontaneous polarization of the single-crystal film throughout the device.[22] Alternatively, this lateral clamping can be used to engineer the in-plane strain response of the PMN-PT membrane devices. Our previous study found this clamping to be independent of electrode size, and allows for control of the in-plane strain anisotropy in a manner which strongly depends on electrode shape and aspect ratio.[23] Through this strain-engineering approach, the PMN-PT membranes can be used in a wide variety of applications, such as low-voltage magnetoelectric coupling for next generation sensing and memory storage technologies.[24]

In summary, using XRD and phase-field simulations, we demonstrate that single-crystal PMN-PT thin films can exhibit bulk-like piezo-behavior once removed from the constraints of substrate clamping. The clamped PMN-PT films exhibit a limited response consisting of reversible volume exchange between the R and $M_a$ phases, while the free PMN-PT membrane exhibits a full rotation between the R <111> and T <001> polarization directions. In addition to the T phase being energetically unfavorable due to tensile strain, the substrate constrains the polarization within the (110) ($M_a$) plane, inhibiting full transition to the T <001>. Releasing the film allows the polarization to access the additional monoclinic orientations ($M_b$ and $M_c$), facilitating the complete transition through the MPB. The impinging strain fields between active and inactive material further reduce the piezoresponse by inhibiting the exchange of R and M/T domains. This suggests that reduction of lateral clamping with inactive material will also increase the response. Our study has provided a critical step towards the engineering of highly responsive piezoelectric MEMS and other devices.

**Supplemental Information**
See supplementary information for PE loops of both clamped and membrane PMN-PT thin films, representative XRD scans and fits for the clamped PMN-PT film, XRD results for clamped film under varying AC bias magnitudes, estimated longitudinal coefficient $d_{33}$ of the clamped film from XRD, thermodynamic calculations of R/M/T free energies in clamped and unclamped PMN-PT films, and supplementary notes 1-4.

**Acknowledgement**
This work was supported by the Army Research Office through grant W911NF-17-1-0462, Vannevar Bush Faculty Fellowship (N00014-20-1-2844), the Gordon and Betty Moore Foundation's EPiQS Initiative, grant GBMF9065 to C.-B.E., AFOSR (FA9550-15-1-0334 (CBE)), and NSF through the University of Wisconsin MRSEC (DMR-1720415 (CBE). Piezoelectric response measurements by double-beam laser interferometry and analysis at the University of Wisconsin–Madison was supported by the US Department of Energy (DOE), Office of Science, Office of Basic Energy Sciences (BES), under award number DE-FG02-06ER46327 (C.B.E.). A.B acknowledge the support from the National Science Foundation Graduate Research Fellowship under Grant No. DGE-1256259. Use of the Advanced Photon Source, an Office of Science User Facility operated for the U.S. DOE Office of Science by Argonne National Laboratory (ANL), was supported by the U.S. DOE under Contract No. DE-AC02-06CH11357. The work at Cornell was made possible by the support from Samsung Electronics Company and the Gordon and Betty Moore Foundation's EPiQS Initiative, Grant GBMF9073 to D.G.S.

**Conflicts of Interest**
The authors have no conflicts to disclose.

**Data Availability**
The data that support the findings of this study are available from the corresponding author upon reasonable request.

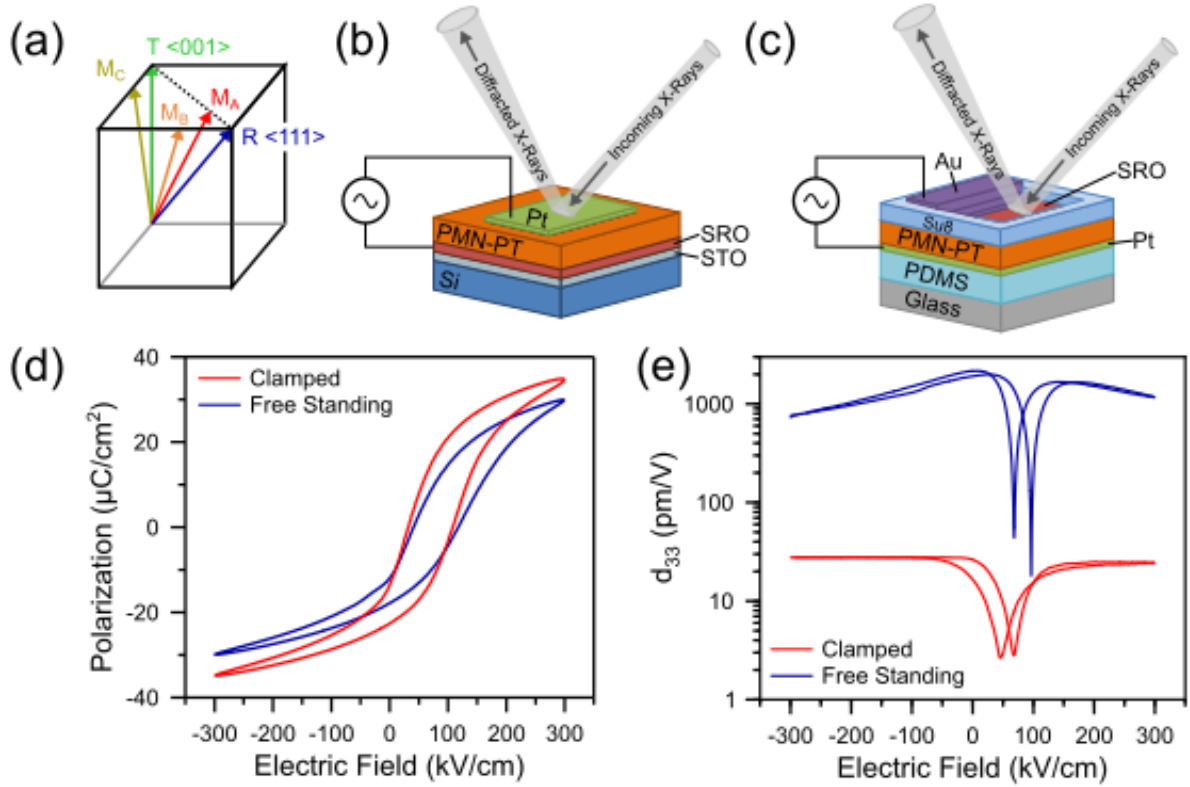

**Figure 1**: Schematic of experimental heterostructures, ferroelectric, and interferometer measurements of clamped and unclamped PMN-PT films. **(a)** Spontaneous polarization directions for R, $M_a$, $M_b$, $M_c$, and T phases. **(b)** Clamped PMN-PT thin film heterostructure. **(c)** Membrane PMN-PT thin film heterostructure. **(d)** PE loops (taken at 10 kHz frequency) and **(e)** double-beam laser interferometry (DBLI) of both clamped and membrane PMN-PT thin films.

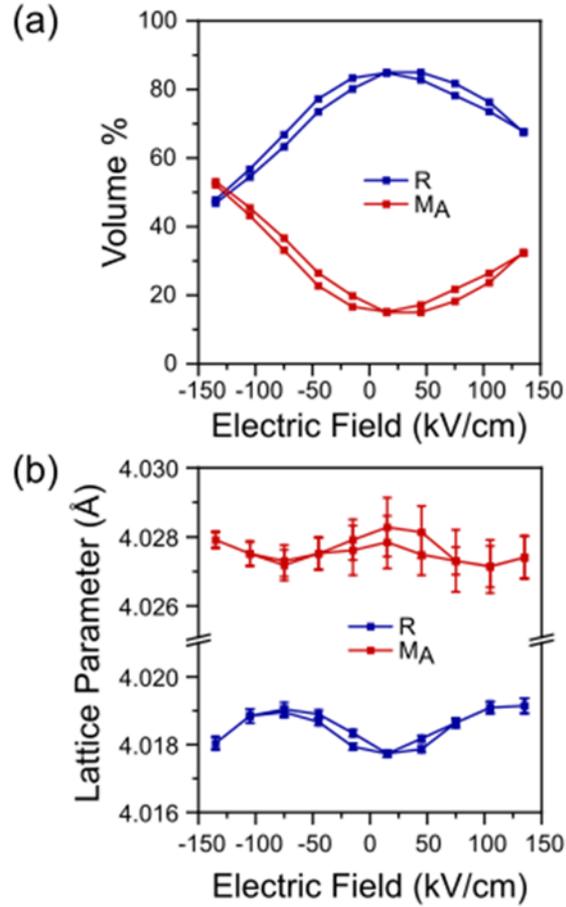

**Figure 2**: XRD data under 150 kV/cm AC bias for the clamped PMN-PT thin film. **(a)** Volume percentage vs electric field of the R and $M_a$ phases present within the clamped PMN-PT film. **(b)** Lattice parameter vs electric field for the R and $M_a$ phases active during AC biasing.

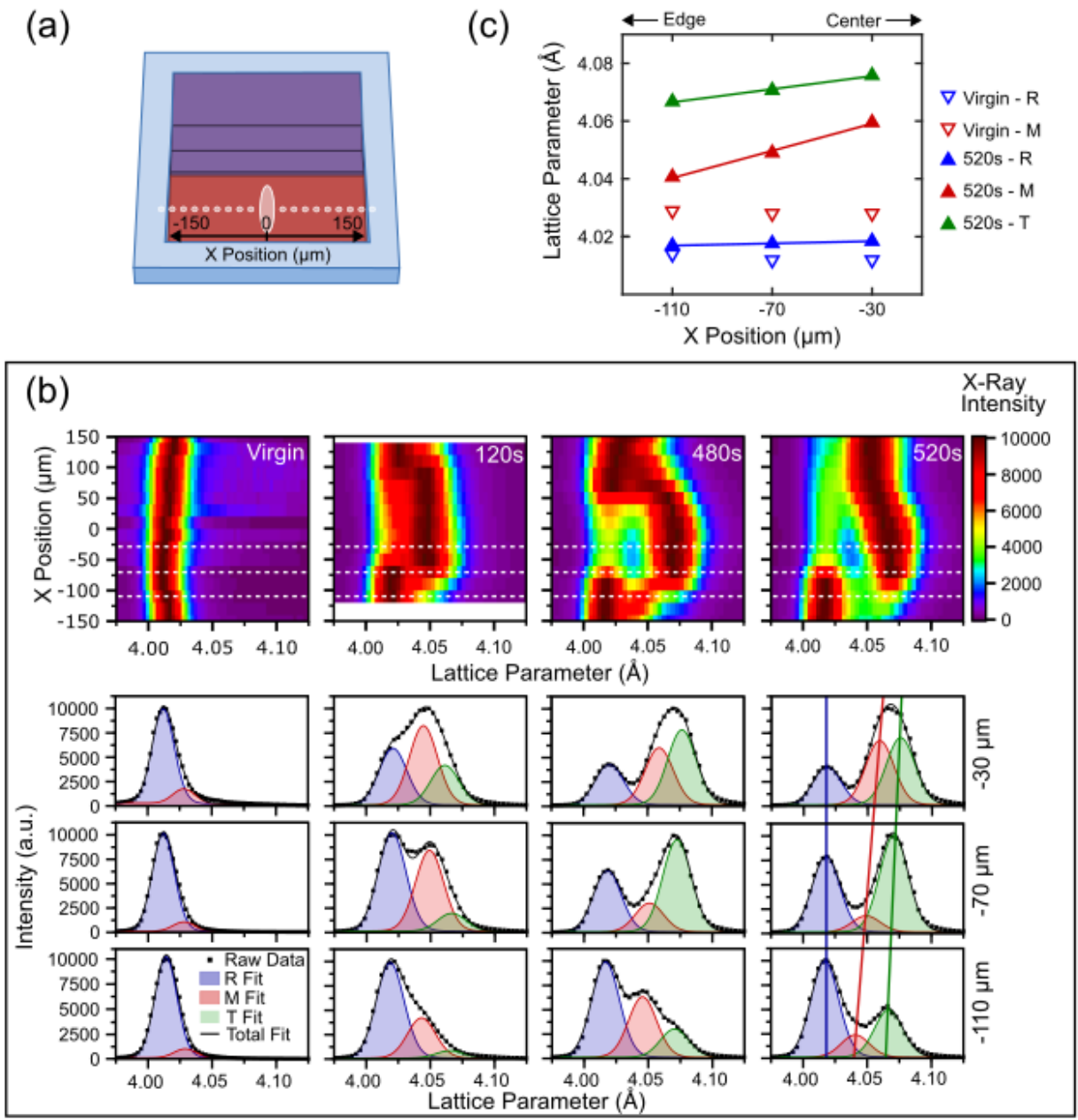

**Figure 3**: Zero-field XRD contrast views of the PMN-PT 004 peak in the membrane after different periods of AC cycling. **(a)** The beam was stepped across the electrode to observe the spatial dependence of the piezoresponse. **(b)** The top panels show the diffraction intensity (colorbar) vs lattice parameter (x-axis) at different locations of the electrode (y-axis) after periods of AC cycling (virgin 0 s, 120 s, 480 s, 520 s). Slices along the positions -110 µm, -70 µm, and -30 µm are shown in the plots below each of the top panels. **(c)** Lattice parameter vs electrode position for the virgin and 520 s states.

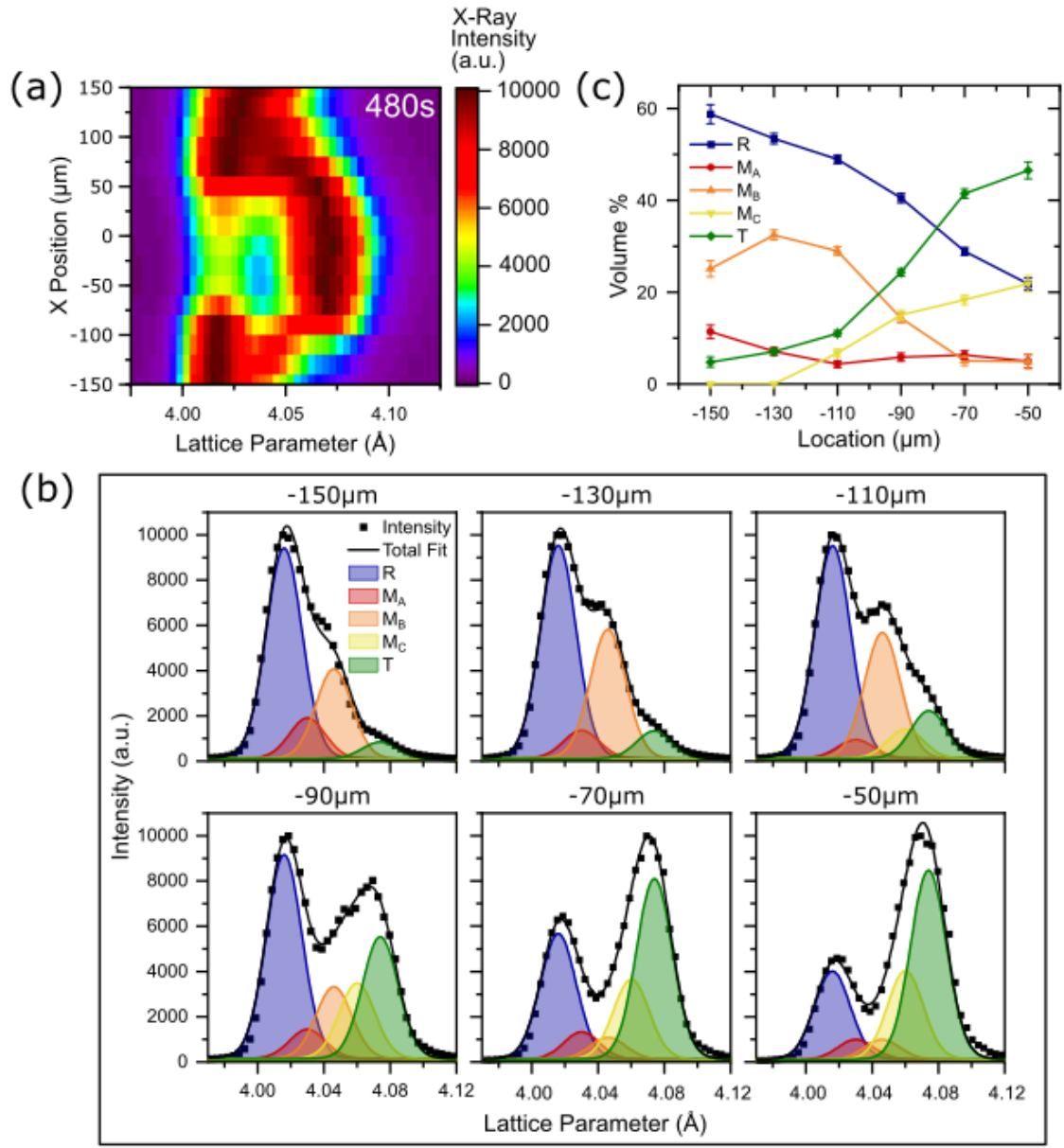

**Figure 4**: Polarization pathway through the MPB. $M_a$, $M_b$, and $M_c$ phase fitting of the 480 s AC field sweep data. **(a)** Diffraction intensity (colorbar) vs lattice parameter (x-axis) as the beam is stepped across the electrode (y-axis) for the 480 s membrane state. **(b)** Slices of the XRD from (a) showing the peak fitting results at different regions of the electrode including all three monoclinic variants. **(c)** Volume percentage of the R, $M_a$, $M_b$, $M_c$, and T phases for fitted locations.

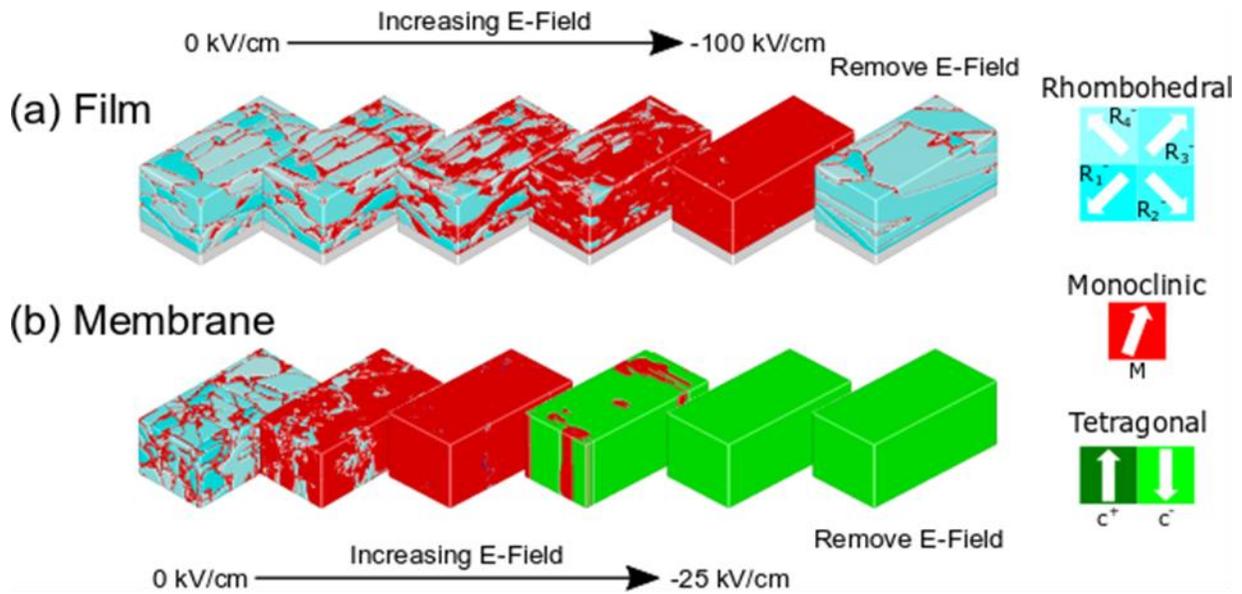

**Figure 5**: Phase-field simulations of both clamped and unclamped PMN-PT films with applied electric fields. **(a)** Evolution of the clamped film's microstructure under an electric field shows the 85% R film becoming ~100% M at -100 kV/cm. Upon field removal, the clamped film becomes primarily R again demonstrating phase reversibility. **(b)** The as-grown microstructure of the membrane remains primarily R after substrate removal. By -25 kV/cm, the film becomes entirely T and remains in the T phase once the field is removed.

Supplemental Information for

# Microscopic piezoelectric behavior of clamped and membrane (001) PMN-30PT thin films

A. Brewer[1]†, S. Lindemann[1]†, B. Wang[2], W. Maeng[1], J. Frederick[1], F. Li[3], Y. Choi[4], P. J. Thompson[5], J. W. Kim[4], T. Mooney[4], V. Vaithyanathan[6], D. G. Schlom[6,7,8], M. S. Rzchowski[9], L. Q. Chen[2], P. J. Ryan[4,10]*, and C. B. Eom[1]*

[1]Department of Materials Science and Engineering, University of Wisconsin-Madison, Madison, WI 53706, USA

[2]Department of Materials Science and Engineering, The Pennsylvania State University, University Park, PA 16802, USA

[3]Electronic Materials Research Laboratory, Key Laboratory of the Ministry of Education, Xi'an Jiaotong University, Xi'an, 710049, China

[4]X-Ray Science Division, Argonne National Laboratory, Argonne, IL 60439, USA

[5]Department of Physics, University of Liverpool, Liverpool, L69 3BX, UK

[6]Department of Materials Science and Engineering, Cornell University, Ithaca, NY 14853, USA

[7]Kavli Institute at Cornell for Nanoscale Science, Ithaca, NY 14853, USA

[8]Leibniz-Institut für Kristallzüchtung, Max-Born-Straße 2, 12489 Berlin, Germany

[9]Department of Physics, University of Wisconsin-Madison, Madison, WI 53706, USA

[10]School of Physical Sciences, Dublin City University – Dublin 9, Ireland

† These authors contributed equally
*Correspondence should be addressed: eom@engr.wisc.edu, pryan@aps.anl.gov

# Supplemental Figures

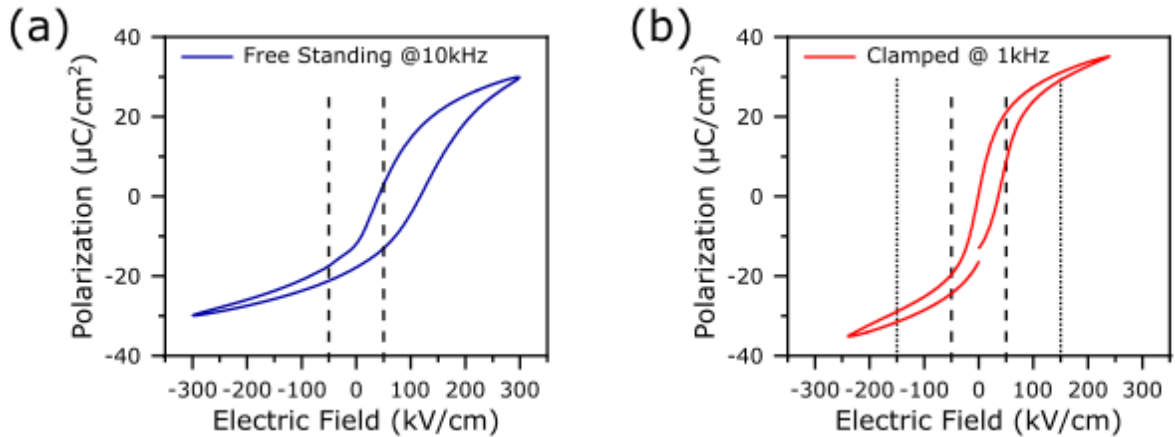

**Fig. S1**: Polarization vs Electric Field loops of the clamped and free-standing membrane PMN-PT films. **(a)** Due to the FE imprint, the AC 50 kV/cm field (dashed lines) drives a one-way evolution through the MPB without providing sufficient energy to switch the polarization. **(b)** PE loop of the clamped film taken at 1kHz frequency, with dashed and dotted lines for the 50 and 150 kV/cm AC fields, respectively, used in the XRD measurements of **Fig. 2** and **Fig. S3**

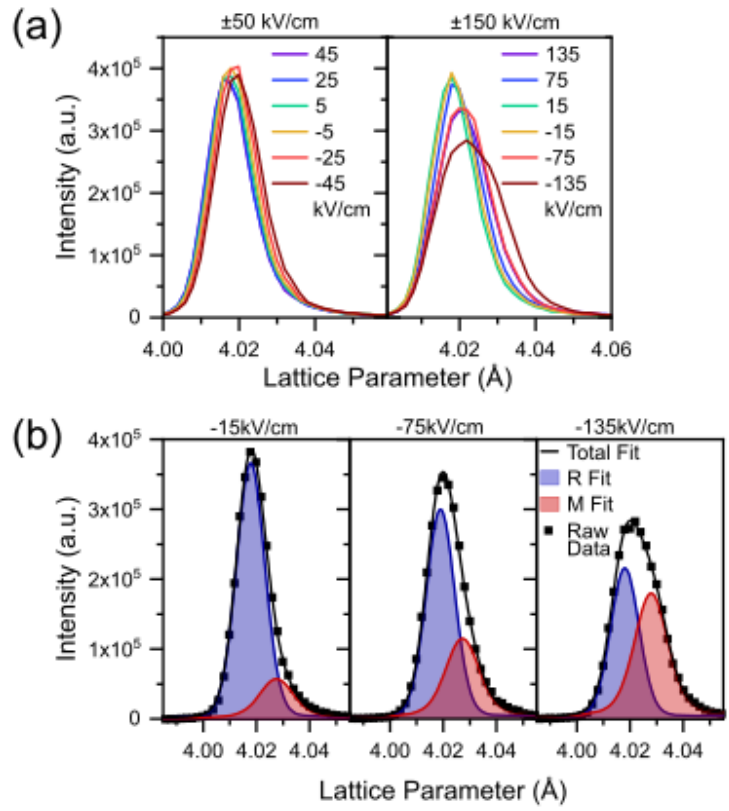

**Fig. S2:** Representative XRD scans for the clamped PMN-PT film with example fittings for selected binned voltages. **(a)** XRD scans of the PMN-PT (004) peak of the clamped film under differing AC bias conditions. **(b)** Fitting of R and Ma phases for selected L-scans of the 150 kV/cm sweep.

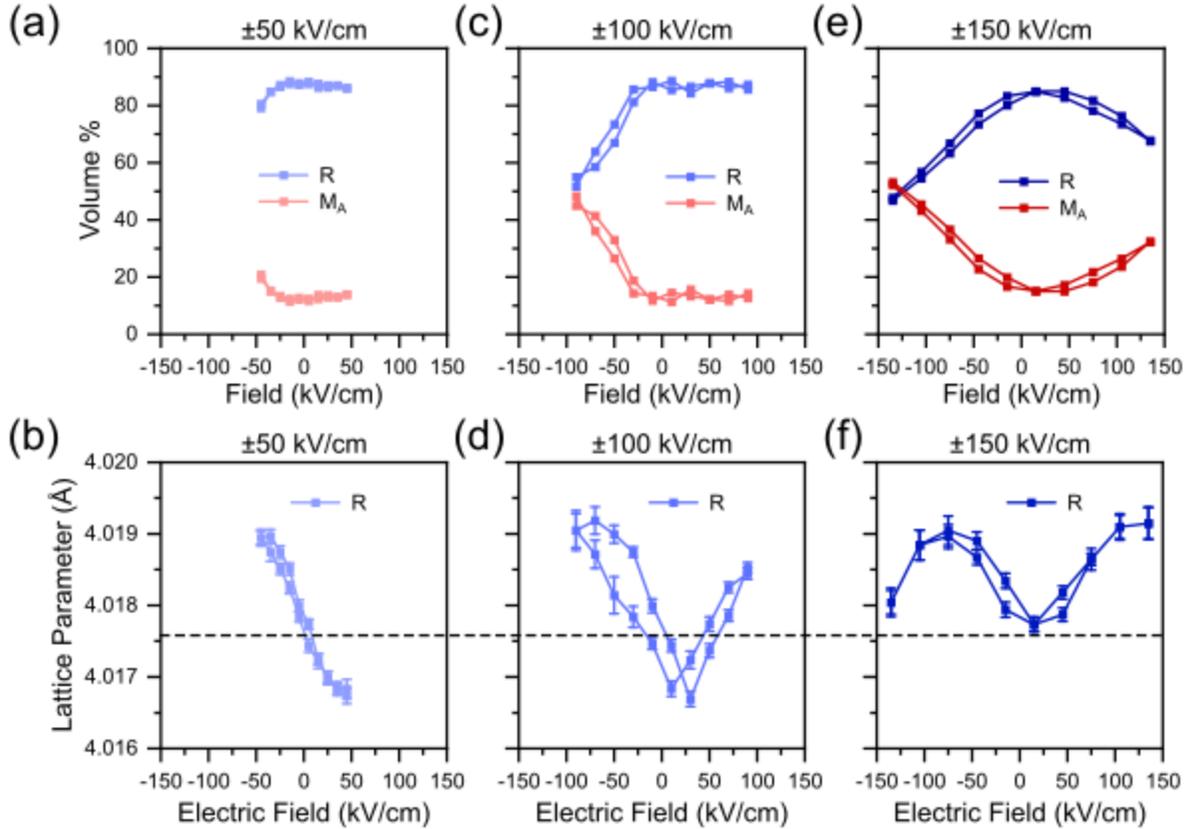

**Fig. S3:** Volume percentage and lattice parameters of the clamped film for different AC field values. **(a)-(b)** At 50 kV/cm, there is minimal volume exchange between R and $M_A$. The R phase exhibits a linear change in lattice parameter with the applied bias. **(c)-(d)** For 100 kV/cm, there is volume exchange between the R and $M_A$ symmetries at negative fields. The R phase exhibits butterfly loop lattice expansion. The asymmetry between positive and negative bias is due to the FE imprint of the PMN-PT film (**Figure S1a**). **(e)-(f)** At 150 kV/cm, volume exchange between R and $M_A$ is present for both positive and negative biases. The dashed line shows that at 150 kV/cm AC, the R phase does not exhibit contraction before switching.

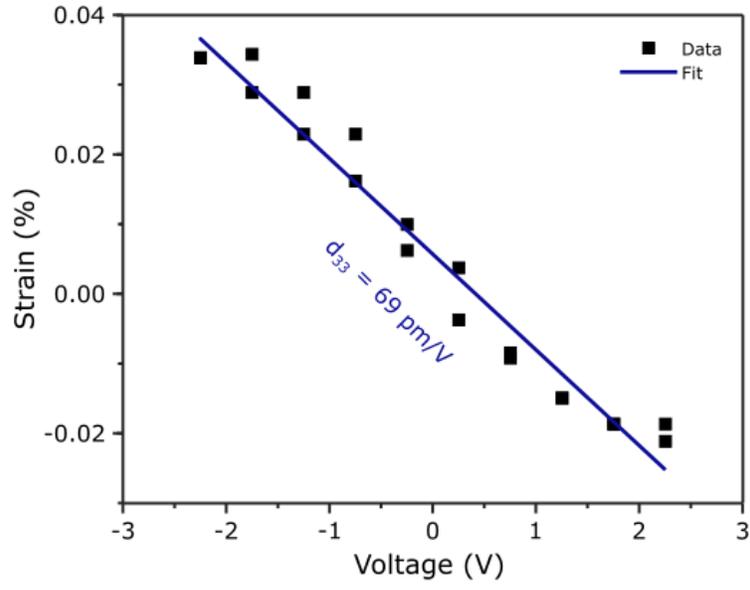

**Fig. S4:** Longitudinal piezoelectric coefficient $d_{33}$ of the clamped PMN-PT film's rhombohedral peak position of the 50 kV/cm sweep in **Fig. 2(c)**. The strain was calculated relative to the lattice parameter at 0 V. The clamped film R phase exhibits an intrinsic $d_{33}$ = 69 pm/V.

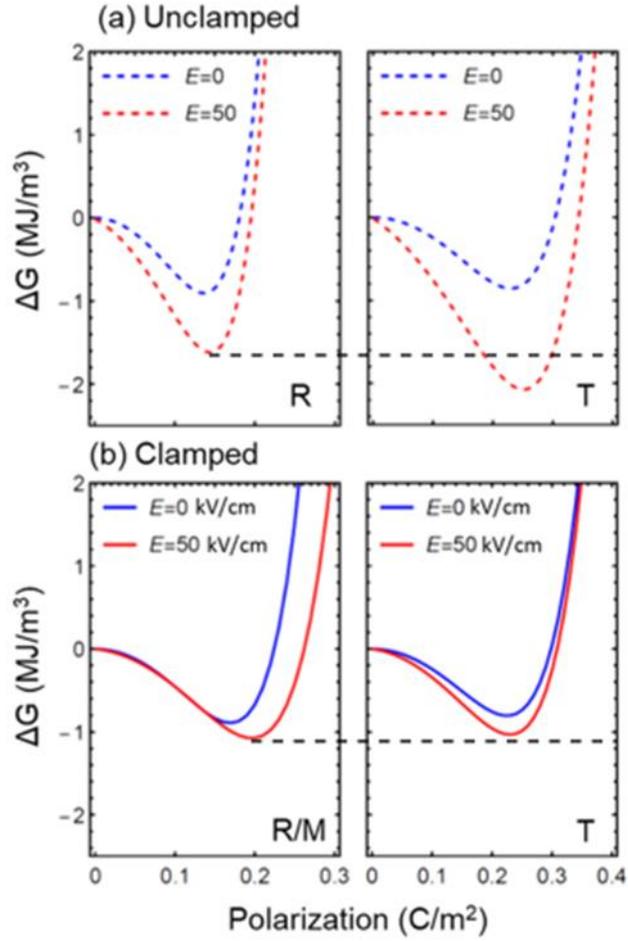

**Fig. S5:** Calculated Gibbs free energy for an unclamped **(a)** and clamped **(b)** PMN-PT thin film with a built-in bias (FE imprint). In both unclamped and clamped cases, the R phase has a lower minimum of free energy without a built-in bias (E = 0 kV/cm). With a 50 kV/cm built-in bias, the T phase exhibits a lower minimum free energy in the unclamped case, while in the clamped film the R phase retains the lower minimum. Therefore, the presence of the built-in bias due to the FE imprint helps stabilize T domains that are nucleated/grown during AC cycling of the PMN-PT membrane (**Fig. 3**).

# Supplemental Notes

## Note 1: Thin film and membrane sample preparation

Fabrication of both clamped and unclamped PMN-30PT thin films begins with a (001)-oriented Si substrate with a 4 degree miscut towards <110>.[13] A SrTiO$_3$ buffer layer of 20 nm is grown on the Si via molecular-beam epitaxy, followed by deposition of SrRuO$_3$ (SRO) (100 nm) and PMN-PT (500 nm) with RF magnetron sputtering. For the clamped film, 100 nm Pt electrodes are deposited and patterned via DC magnetron sputtering and lift-off photolithography, completing fabrication of the clamped film heterostructure. This patterning defines a boundary between active (biased) and inactive (unbiased) PMN-PT. For the membrane, 100 nm of continuous Pt is deposited on the PMN-PT to act as the bottom electrode in the final heterostructure. Polydimethylsiloxane (PDMS) was spin-coated onto glass and the sample is placed Pt-side down onto the PDMS. The Si substrate was removed by wet etching using a 1:1:1 ratio of HF:HNO$_3$:H$_2$O etchant. The exposed SRO was patterned by photolithography and wet etching using a 4 M solution of NaIO$_4$, followed by addition of the Su8 (3 μm) protective layer via photolithography. Finally, offset Au electrodes (30 nm) are deposited by DC magnetron sputtering and patterned with photolithography and wet etching.

## Note 2: AC X-Ray diffraction methodology

XRD measurements with *in-situ* AC biasing were performed at the Advanced Photon Source Beamline 6-ID-B. 10 keV x-rays were used with a 20 μm x 34 μm beam spot size on the sample, allowing us to scan across the biased region of the 200 μm x 300 μm membrane device. Using a triangular wave signal for the AC sweeps, and custom electronics to synchronize between the x-ray detector and the AC sample bias, a binning procedure that averaged detector counts within specified voltage ranges allowed a measure of the diffraction with applied voltage. For the membrane, 50 kV/cm AC cycling at 1kHz frequency was performed for periods of 0s, 120s, 480s, 520s which correspond to 0, 120k, 480k, and 520k total number of E-field cycles, respectively. The minor loops used in the membrane XRD experiment did not allow for complete switching of DOWN domains to UP, therefore, no signs of ferroelectric fatigue were observed. The lattice parameter and relative phase volume vs electric field was extracted by fitting the L-scans of each voltage bin with a Voigt function.[14] Each phase is subsequently identified by lattice length in comparison to bulk taking into consideration in-plane strain.[15,16] The SrRuO$_3$ film peak has a lattice parameter of 3.932Å. The XRD scans were performed from 3.975Å to 4.125Å, meaning that the SRO peak was below the range of the scans of Figure 3 and Figure 4.

## Note 3: Thermodynamic calculation and phase-field simulation

The Landau-Ginzburg-Devonshire theory of ferroelectrics is used to perform the thermodynamic analysis for PMN-PT clamped and freestanding thin films. The Landau parameters and elastic and electrostrictive coefficients are adopted from literature.[19] In the thermodynamic analysis, the clamped film is modeled as a single-domain state subject to in-plane constraints and out-of-plane stress free condition, following [Ref. 20], while the freestanding film is modeled as a single-domain stress-free bulk state. The different ferroelectric phases are determined by identifying the angle of the polarization with respect to the high-symmetry directions. A threshold angle of 10 degrees is set to distinguish R, M, and T phases, following [Ref. 21]. For example, if the angle of the vector with respect to [111] direction is less than 10 degrees, it is regarded as the R phase. If it deviates more than 10 degrees towards the [001] direction, it is regarded as M until the angle between the polarization vector and [001] is less than 10 degrees, then it is labelled as T phase.

The phase-field simulation[18] is performed to obtain the domain structures for PMN-32PT clamped and freestanding thin films at equilibrium subject to built-in and external electric field. The same parameters used in the thermodynamic calculation are adopted for the phase-field model. The three-dimensional system is discretized into 128×64×64 grid points with each grid representing 1 nm in length. The film thickness is set to be 50 nm. For the clamped film case, the mixed-type mechanical boundary condition is adopted with a substrate layer of 10 nm to relax the mechanical displacement. For the free-standing film, the rest grid points along the out-of-plane directions are treated as buffer layers. For both cases, the short-circuit boundary conditions are adopted on the top and bottom of the film layer while a downward built-in field of 50 kV/cm is imposed. All simulations start from a random noise distribution of polarization vectors, and sufficient simulation timesteps are used to allow the domain structure to evolve into the equilibrium states.

## Note 4: Calculation of theoretical $d_{33}$ in PMN-30PT membranes

For the PMN-30PT membrane, using XRD we found that there is an R phase with lattice parameter of 4.018Å, and a T phase with 4.070Å. The amount of out-of-plane expansion (strain) in one unit cell would therefore be

$$\varepsilon = \frac{\Delta l}{l_0} = \frac{4.070\text{Å} - 4.018\text{Å}}{4.018\text{Å}} = 0.0129 \text{ or } 1.29\%$$

For a 500nm film, if we assume that all of the unit cells would transition from R to T under 5V applied bias (the max field used on the membrane in this study was +/- 50 kV/cm, i.e. +/- 2.5V for the 500nm thick membrane), then the total expansion for the 500nm thick PMN-PT membrane would be:

$$\Delta l = \varepsilon * l_0 = .0129 * 500nm = 6.47nm$$

Therefore, the estimated $d_{33}$ would be:

$$d_{33} = \frac{\varepsilon_3}{E_3} = \frac{\Delta l}{l_0} * \frac{l_0}{V} = \frac{\Delta l}{V} = \frac{6.47nm}{5V} = 1.29\frac{nm}{V} \text{ or } 1290\frac{pm}{V}$$

The estimated $d_{33}$ of 1300 pm/V is consistent with our DBLI measured 1000 pm/V. The difference could arise due to the fact that we may not have 100% conversion of R to T domains in the PMN-PT membranes.